\documentclass{ws-ijmpa}
\usepackage[numbers,sort&compress]{natbib}
\usepackage[english]{babel}
\begin{document}
\markboth{Authors' Names}
{Instructions for Typing Manuscripts (Paper's Title)}
%
\catchline{}{}{}{}{}
%
\title{STUDIES OF PROPERTIES OF THE $\eta'$ MESON\\AT THE COSY--11 FACILITY
}
\author{E.~CZERWI\'NSKI\footnote{e-mail address:
eryk.czerwinski@lnf.infn.it}~$^{1,2,3}$, J.~KLAJA$^{1,2}$, P.~KLAJA$^{2,4}$, and P.~MOSKAL$^{1,2}$
}
\address{
$^1$~Institute of Physics, Jagiellonian University, PL-30-059 Cracow, Poland\\
$^2$~Institute for Nuclear Physics and J\"ulich Center for Hadron Physics,\\Research Center J\"ulich, D-52425 J\"ulich, Germany\\
$^3$~INFN, Laboratori Nazionali di Frascati, IT-00044 Frascati, Italy\\
$^4$~Physikalisches Institut, Friedrich-Alexander-Universit\"at Erlangen-N\"urnberg,\\D-91058 Erlangen, Germany\\
}
\maketitle
\begin{history}
\received{Day Month Year}
\revised{Day Month Year}
\end{history}
\begin{abstract}
We shortly discuss
results on the isospin dependence of the $\eta'$ production cross section
in nucleon-nucleon collisions, 
results of comparative analysis of the invariant mass distributions for the $pp\to pp\eta'$
and $pp\to pp\eta$ reactions,
and we present the value of the total width of the $\eta'$ meson as derived directly
from the mass distribution measured with the mass resolution of 0.3~MeV/c$^2$.
\end{abstract}
\ccode{PACS numbers: 3.60.Le, 13.60, 13.75.Cs, 13.75.-n, 13.85.Lg, 14.20.Dh, 14.40.Aq, 14.40.-n}

The COSY--11 detector 
permits to determine momentum vectors of outgoing nucleons and the four-momentum of unregistered 
meson is reconstructed via the missing mass technique.
The detector was described in many publications and therefore
for detailed information the interested
reader is referred e.g. to articles~\cite{Brauksiepe,AIP,hab}. 

The upper limit of the total cross section for quasi-free $pn \to pn\eta^{\prime}$
reaction has been determined in the excess energy range near the kinematic threshold~\cite{jklaja-phd,jklaja_prc}.
The measurement has been carried out 
using a proton beam and a deuteron cluster target.
The energy dependence of the upper limit of the cross section is extracted with a fixed
proton beam momentum of $p_{beam}=3.35$~GeV/c and exploiting the Fermi
momenta of nucleons inside the deuteron~\cite{hab}. 
The data cover a range of centre-of-mass
excess energies from 0 to 24~MeV.
The experimentally determined upper limit of the ratio
$R_{\eta^{\prime}}~=~{{\sigma(pn \to pn\eta^{\prime})} \over {\sigma(pp \to pp\eta^{\prime})}}$,
which is smaller than the ratio for the $\eta$ meson~\cite{moskal-prc79}, excludes the excitation of the
S$_{11}$(1535) resonance as a dominant production mechanism of the $\eta^{\prime}$ meson in
nucleon-nucleon collisions~\cite{cao-prc781}.
At the same time, the determined upper limits of R$_{\eta^{\prime}}$ go in the direction
of what one would expect in the glue production and production via mesonic currents~\cite{bass,kampfer-ep1}.
For quantitative tests of these mechanisms an order of magnitude larger statistics and
a larger energy range would be required. This can be reached with the WASA-at-COSY facility~\cite{wasa}.
The detailed description of the data evaluation as well as theoretical motivation one can find in Ref.~\cite{jklaja-phd,jklaja_prc,moskal-prc79,jpg2006}

The COSY-11 collaboration measured also the $pp~\to~pp\eta$ and
$pp~\to~pp\eta^{\prime}$ reactions in order to perform comparative studies of the
interactions within the proton-proton-meson system~\cite{hab}.
The experiment results in differential distributions of squared invariant
proton-proton ($s_{pp}$) and proton-$\eta^{\prime}$ ($s_{p\eta^{\prime}}$) masses, as well as in
angular distributions and the total cross section at an excess energy of 16.4~MeV~\cite{pk_phd,plb_klajus}.
The differential distributions $s_{pp}$ and $s_{p\eta^{\prime}}$ are compared to theoretical
predictions~\cite{kanzo,deloff} and to the analogous spectra determined for the $pp \to pp\eta$ reaction~\cite{prc69}.
The comparison of the results for the $\eta$ and $\eta^{\prime}$ meson production
rather excludes the hypothesis that the enhancement observed in the invariant mass distributions
is due to the meson-proton interaction. Further, the shapes of the distributions do not favour any of the
postulated theoretical models.

The reaction $pp~\to~pp\eta^{\prime}$ measured at five different beam momenta was used
for the determination of the total width of the $\eta'$ meson, which
was directly derived from the mass distributions~\cite{ErykPhD,ErykQCD}.
Based on the  $\eta'$ meson mass spectra reconstructed for  2300 events with the experimental resolution 
of 0.3~MeV/c$^2$ 
the total width of the $\eta'$ meson was determined  to be
$\Gamma_{\eta'}=0.226\pm0.017(\textrm{stat.})\pm0.014(\textrm{syst.})$~MeV/c$^2$,
which is the most precise measurement until now.


\begin{thebibliography}{00}
\bibitem{Brauksiepe} S.~Brauksiepe {\it et~al.}, Nucl. Instr. Meth. A \textbf{376}, 397 (1996).
\bibitem{AIP}P.~Klaja {\it et al.}, {\it AIP Conf. Proc.} {\bf 796}, 160 (2005).
\bibitem{hab} P. Moskal  e-Print Archive: hep-ph/0408162, (2004).
\bibitem{jklaja-phd} J.~Klaja,   PhD thesis, arXiv:~0909.4399, (2009).
\bibitem{jklaja_prc} J.~Klaja {\it et al.}, Phys. Rev. {\bf C 81}, 035209 (2010).
\bibitem{moskal-prc79} P.~Moskal {\it et al.,} Phys. Rev. {\bf C 79}, 015208 (2009).
\bibitem{cao-prc781} Xu~Cao and Xi-Guo~Lee, Phys. Rev. {\bf C 78}, 035207 (2008).
\bibitem{bass}S.~D.~Bass, Phys. Lett. {\bf B 463}, 286 (1999).
\bibitem{kampfer-ep1} L.~P.~Kaptari, B.~K{\"a}mpfer, Eur. Phys. J. {\bf A 37}, 69 (2008).
\bibitem{wasa} H.-H. Adam {\it et al.},e-Print: nucl-ex/0411038 (2004).
\bibitem{jpg2006} P.~Moskal {\it et al.,} J. Phys. {\bf G 32}, 629 (2006).
\bibitem{pk_phd} P. Klaja, PhD thesis, arXiv:~0907.1491 (2009).
\bibitem{plb_klajus} P. Klaja {\it et al.}, Phys. Lett. {\bf B~684}, 11 (2010).
\bibitem{deloff} A. Deloff, {Phys. Rev. \bf C~69}, 035206 (2004).
\bibitem{kanzo} K. Nakayama {\it et al.}, Phys. Rev. {\bf C~68}, 045201 (2003).
\bibitem{prc69} P. Moskal {\it et al.}, Phys. Rev. {\bf C~69}, 025203 (2004).
\bibitem{ErykPhD} E.~Czerwi\'nski, PhD thesis, arXiv:~0909.2781, (2009).
\bibitem{ErykQCD} E.~Czerwi\'nski {\it et al.}, arXiv:1007.2277 (2010).
\end{thebibliography}
\end{document}